\documentclass[twocolumn,prl,superscriptaddress]{revtex4-2}
\usepackage{graphics}
\usepackage{graphicx}
\usepackage{amsfonts}
\usepackage{textcomp}
\usepackage{amssymb}
\usepackage{mathrsfs}
\usepackage{amsmath}
\usepackage{color}
\usepackage{float}

\begin{document}

\title{Emergence of Disordered Hyperuniformity in Melts of Linear Diblock Copolymers}

\author{Duyu Chen}
\email[correspondence sent to: ]{duyu@alumni.princeton.edu}
\affiliation{Materials Research Laboratory, University of California, Santa Barbara, California 93106, United States}
\author{Michael A. Klatt}
\affiliation{Institut für KI Sicherheit, Deutsches Zentrum für Luft‐ und Raumfahrt (DLR), Wilhelm‐Runge‐Straße 10, 89081 Ulm, Germany;
Institut für Materialphysik im Weltraum, Deutsches Zentrum für Luft- und Raumfahrt (DLR), 51170 Köln, Germany;
Department of Physics, Ludwig-Maximilians-Universität, Schellingstraße 4, 80799 Munich, Germany}
\author{Glenn H. Fredrickson}
\email[correspondence sent to: ]{ghf@mrl.ucsb.edu}
\affiliation{Materials Research Laboratory, University of California, Santa Barbara, California 93106, United States} \affiliation{Department of Chemical Engineering,
University of California, Santa Barbara, California 93106,
United States}


\begin{abstract}

Disordered hyperuniform (DHU) systems are recently discovered exotic states of matter, where (normalized) large-scale density fluctuations are completely suppressed as in crystals, even though the systems are isotropic and lack conventional long-range order. Despite recent success, realizing such systems using bottom-up approaches remains challenging. Here, we study the large-scale behavior of neat melts of linear diblock copolymers using large-cell self-consistent field theory (SCFT) simulations. We initialize SCFT simulations using point patterns that correspond to the local energy minimum of the so-called Quantizer energy, a geometric functional related to the free energy of copolymeric self-assemblies. Upon relaxation via the SCFT simulations, we obtain a new class of metastable disordered micelle mesophases that are hyperuniform. Moreover, we find that DHU micelle mesophases possess very similar free energies to the thermodynamically stable body-centered cubic sphere mesophases and are also much more favorable energetically than previously obtained liquid-like packings. Our findings shed light on the design of novel disordered hyperuniform materials using bottom-up approaches, and suggest new possibilities for technological applications, e.g., novel non-iridescent structural colors.






\end{abstract}
\maketitle

{\it Disordered hyperuniform} (DHU) systems are recently discovered exotic states of matter \cite{To03, To18a} that combine properties of both perfect crystals and liquids. These systems are similar to liquids or glasses in that they are statistically isotropic and possess no Bragg peaks, i.e., they lack conventional long-range translational and orientational order; nevertheless, they completely suppress (normalized) large-scale density fluctuations, like crystals, and in this sense, they possess a \textit{hidden} long-range order \cite{To03, Za09, To18a}. In particular, DHU many-body systems in $d$-dimensional Euclidean space $\mathbb{R}^d$ are characterized by a vanishing structure factor $S(k)$ in the infinite-wavelength limit, i.e., $\lim_{k\rightarrow 0}S(k) = 0$, where $k$ is the wavenumber \cite{To03, To18a}. Equivalently, these systems can be characterized by a vanishing normalized local number variance $\sigma^2(R)/R^d$ in the large-$R$ limit, where $\sigma_N^2(R)\equiv \langle N^2(R)\rangle - \langle N(R) \rangle^2$ is the local number variance associated with a spherical window of radius $R$, $N(R)$ denotes the number of particles in a spherical window of radius $R$ randomly placed into the system, and $\langle \cdots \rangle$ denotes an ensemble average \cite{To03, To18a}. DHU states have been discovered in a variety of equilibrium and non-equilibrium physical and biological systems, and their exotic structural features have been shown to endow such systems with desirable physical properties that cannot be achieved in either ordinary disordered or perfectly crystalline states \cite{To03, To18a, Fl09, Ma13, Ji14, Dr15, He15, Ja15, Zh16, Ru19, Hu21, Le19a, Le19b, Xu17}. For example, certain DHU dielectric networks \cite{Fl09, Kl22} were found to possess complete photonic band gaps comparable in size to photonic crystals, while at the same time maintaining statistical isotropy, enabling waveguide geometries not possible with photonic crystals \cite{Ma13}. Today, most DHU materials are designed ``top-down'' since a realization via a ``bottom-up'' self-assembly remains challenging~\cite{To18a}, despite a few such examples~\cite{Ch21a, Sa20a} .

Recently, Chremos and coworkers \cite{Ch18b} have found that the cores of stars and the backbones of bottlebrush polymers possess a hyperuniform distribution in their melts, although the mesophases formed by these stars and bottlebrushes overall are not hyperuniform. In a recent work \cite{Kl19}, some of us have demonstrated that the local potential-energy minima associated with the so-called ``Quantizer energy'' are effectively hyperuniform. Interestingly, this Quantizer energy is reminiscent of the entropic chain stretching free energy of copolymeric self-assemblies \cite{Gr03,Re18}. For example, in the diblock foam model of block copolymer melts \cite{Re18}, the free energy is expressed as essentially the weighted sum of an interfacial energy term and the Quantizer energy. These earlier results motivate the search for disordered hyperuniform mesophases in block copolymer systems.


While the equilibrium mesophases of block copolymers are well studied by theory and experiment \cite{Ba90, Ba99}, the self-assembly of block copolymers has sometimes been observed to suffer from slow dynamics, making it difficult to achieve the true equilibrium state \cite{Ba17, Lo22}. In particular, near the order--disorder transition (ODT), upon rapid cooling from the disordered homogeneous state, systems are often kinetically trapped into a mysterious ``liquid-like'' packing (LLP) state, whose scattering pattern typically exhibits a broad peak and lacks the sharp Bragg peaks associated with ordered morphologies \cite{Ki17, Ki18, Gi16, Ba20a}. This LLP state also exhibits essentially solid-like mechanical properties while maintaining a disordered morphology, and thus may be thought of as a soft glass \cite{Gi16}. However, it is currently unclear from the available experimental data whether these disordered packings are hyperuniform or non-hyperuniform. Recently, using large-cell self-consistent field theory (SCFT) simulations initialized from the structure of a Lennard-Jones fluid, Dorfman and Wang \cite{Do23} demonstrated that compositionally asymmetric diblock copolymers possess a rugged free energy landscape with vast numbers of metastable liquid-like states corresponding to distinct aperiodic packings of spherical micelles. They further postulated that this vast population of near-degenerate  liquid-like states is responsible for the slow ordering kinetics. However, given the particular choice of the Lennard-Jones fluid structure as the initial condition for the sphere centers and the fact that the number of spheres employed in the SCFT simulations was relatively small (54 micelles), questions regarding the nature of the disorder and the structural characterization of the liquid-like states remain, in particular at large length scales. We also note that previous SCFT simulations \cite{Ya18} have been used to probe the properties of disordered mesophases in melts of compositionally symmetric linear AB diblock copolymers as well.




Here, we directly and rigorously show, for the first time, that DHU mesophases are energetically competitive in neat melts of block copolymers, and to this end establish a link between the two research fields of disordered hyperuniform systems and block copolymers. In particular, we obtain a new class of metastable mesophases consisting of disordered A-rich micelle cores distributed in a continuous B-rich matrix in neat melts of AB linear diblock copolymers. Specifically, we first develop a Gaussian kernel method to initialize SCFT simulations
of disordered micelles using point patterns that
correspond to the local energy minima of the Quantizer
energy as the micelle centers. To evaluate the large-scale density fluctuations, we then perform large-cell SCFT simulations containing as many as 2000 micelles in the simulation box, and obtain metastable mesophases that are local free energy minima. Importantly, we find that these mesophases are DHU, which distinguishes them from typical liquid or glass morphologies. Moreover, these metastable mesophases possess very similar free energies to the thermodynamically stable BCC sphere phase (with free energy per chain $\langle F_{DHU} \rangle \approx 1.0003\times F_{BCC}$), and are approximately 60\% lower in free energy relative to the BCC sphere phase than previously obtained \cite{Do23} liquid-like packings. Our findings suggest a new bottom-up route to realize DHU materials.


\textit{Initialization of SCFT.}
The micelles formed by block copolymers often contain a large number of polymer chains and each polymer chain typically consists of a large number of atoms. This makes it computationally difficult or intractable to study copolymer mesophases using all-atom or even coarse-grained particle-based simulations over time scales relevant to the self-assembly of block copolymers \cite{Fr06}. Fortunately, exact mathematical transformations exist that can decouple the many-body interactions among polymer chains and replace them with independent interactions between a single polymer chain and one or more auxiliary potential fields $w({\bf r})$ \cite{Fr06}. Among these field-based simulations, SCFT simulations \cite{Fr06} are ones based on a mean-field approximation, and over the past few decades have enjoyed considerable success in predicting the self-assembly behavior of various block copolymer systems [see Supplemental Material (SM) for a brief introduction to SCFT simulations]. In this work, we employ discrete Gaussian chain models for incompressible melts of AB linear diblock copolymers. The segmental interactions are described by a Flory-Huggins interaction parameter $\chi$ that favors similar monomer contacts (A-A and B-B) over dissimilar contacts (A-B). It is well-known at the mean-field (or SCFT) level, the equilibrium phase behavior of conformationally symmetric AB linear diblock copolymers is dictated by the volume fraction $f_A$ of the minority monomer species of A and segregation strength $\chi N$, where $N$ is the degree of polymerization for a polymer chain. On the other hand, the effect of temperature on the phase behavior is taken into account via $\chi$, as $\chi$ typically has temperature dependence $\chi (T)=A/T+B$ with $A>0$ \cite{Fr06}. Here $B$ is a constant and $T$ is temperature. Since the exact value of $\chi$ also depends on the specific types of the monomer species, the dependence of phase behavior on temperature is not universal.

To model disordered micelle mesophases, we initialize the SCFT simulations using different disordered point patterns as initial sphere locations via a Gaussian kernel method that we have developed (see SM for details). In particular, we use disordered point configurations of the local energy minima of the Quantizer problem \cite{To10}. Roughly speaking, the Quantizer problem is defined as the optimization of the moment of inertia of Voronoi cells, i.e., similarly-sized sphere-like polyhedra that tile space are preferred. Specifically, for a given set of $n$ ``generating'' points $\mathbf{z}_1,\cdots,\mathbf{z}_n$ the Voronoi partition is the partition of space into cells, where each Voronoi cell $C_i$ consists of all points in space that are closer to $\mathbf{z}_i$ than to any other point $\mathbf{z}_{j\neq i}$. The ``Quantizer energy'' is the total energy $\sum_{i} E(C_i,\mathbf{z}_i)$ summing up all individual cell-contributions, and the cell energy $E(C_i,\mathbf{z}_i)$ is given by $E(C_i,\mathbf{z}_i) \equiv \int_{C_i} || \mathbf{x} - \mathbf{z}_i ||^2 d \mathbf{x}$. We also consider other disordered point patterns such as Poisson point patterns that are totally random, normal liquids, random close packings of hard spheres, and disordered hyperuniform stealthy point patterns (see SM for details).

\begin{figure}[ht!]
\begin{center}
$\begin{array}{c}\\
\includegraphics[width=0.48\textwidth]{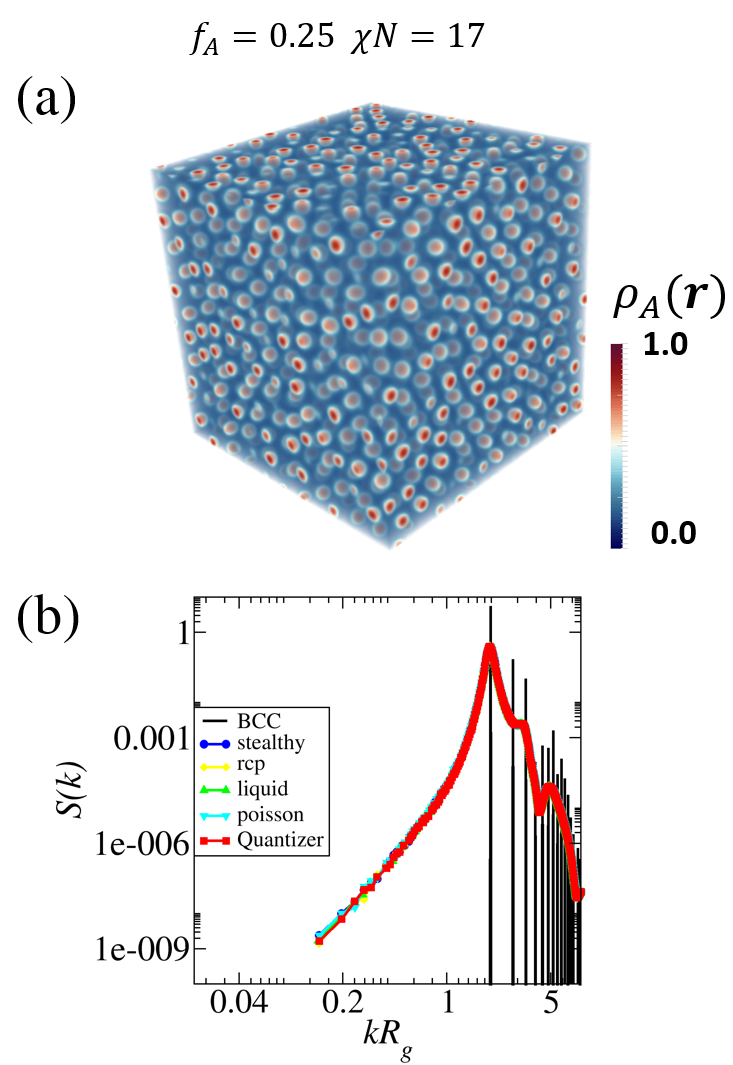} 
\end{array}$
\end{center}
\caption{(Color online) (a) Visualization of the SCFT-solved density field $\rho_A({\bf r})$ of a metastable mesophase consisting of A-rich micelles distributed in B-rich matrix in neat melts of AB linear diblock copolymers at $\chi N = 17$ and $f_A = 0.25$. (b) Ensemble-averaged structure factor $S(k)$ of the disordered micelle mesophases resolved by SCFT simulations at $\chi N = 17$ and $f_A = 0.25$ using different disordered point patterns as initial sphere locations, as well as that of a BCC sphere phase containing 2000 micelles at the same $\chi N$ and $f_A$. We consider disordered point patterns such as point configurations of the local energy minima of the Quantizer problem, Poisson point patterns, liquids, random close packings of hard spheres, and disordered hyperuniform stealthy point patterns (see SM for details).} \label{fig_1}
\end{figure}

\textit{Disordered Hyperuniform Micelles.}
Next, we run large-cell SCFT simulations (see SM for the simulation details) of disordered packings of 2000 A-rich micelles distributed in the B-rich matrix in an initial cell corresponding to $10\times10\times10$ BCC unit cells in size. In our SCFT simulations we employ periodic boundary conditions and maintain the box shape to be cubic, while allowing the side length of the box to vary, which is consistent with the boundary conditions used to generate the initial point patterns. Mathematically, any mesophase formed by incompressible melts of AB linear diblock copolymers is uniquely specified by the density field $\rho_A({\bf r})$ (normalized by the average density $\rho_0$), which is a scalar field with values between 0 and 1. Its scattering intensity is proportional to the spectral density $\tilde{\psi}(k)$ \cite{To16a, Ma17a}, which can be used to analyze the hyperuniformity of the mesophase. The density field $\rho_B({\bf r})$ is related to $\rho_A({\bf r})$ via $\rho_B({\bf r})=1-\rho_A({\bf r})$ due to the incompressible melt condition. In the polymer physics community, this quantity $\tilde{\psi}(k)$ is referred to as the structure factor $S(k)$, and here we adopt this convention. 

In Fig. \ref{fig_1}(a) we visualize the density field $\rho_A(r)$ of a SCFT-optimized mesostructure for a disordered packing of 2000 micelles at $\chi N = 17$ and $f=0.25$, and the ensemble-averaged $S(k)$ [see Supplemental Material for the detail of $S(k)$ calculation] of the mesostructures relaxed by SCFT simulations using the aforementioned wide variety of disordered point patterns as initial micelle centers are shown in Fig. \ref{fig_1}(b), which are averaged over 10 configurations. We find that all these SCFT simulations lead to very similar final relaxed structures with essentially identical pair statistics as shown in Fig. \ref{fig_1}(a) and virtually the same free energies. However, SCFT simulations initialized using the point patterns associated with the local energy minima of the Quantizer problem (henceforth referred to as the Quantizer point patterns) are much quicker to converge than those initialized using the other disordered point patterns, suggesting the Quantizer point patterns are far more realistic than the other disordered point patterns for the micelle centers in the melts of AB linear diblock copolymers. Therefore, we will henceforth use these Quantizer point patterns as initial micelle centers for our SCFT simulations. As a comparison we also include $S(k)$ for a $10\times10\times10$ periodic replica of a BCC sphere mesophase in a conventional cell at the same condition. Interestingly, $S(k)$ of the mesophases consisting of disordered micelles possesses a broad primary peak with no other sharp Bragg peaks, indicating the lack of conventional translational and/or orientational order commonly seen in crystals or quasicrystals. The structure factor $S(k)$ of the BCC sphere, on the other hand, possesses multiple sharp Bragg peaks, as expected. Importantly, $S(k)$ of the disordered micelles is approaching zero as $k$ goes to zero (see inset), i.e., the corresponding mesophases are hyperuniform, analogous to crystals or quasicrystals. This behavior is different from typical liquids or molecular glasses, where one would expect $S(k)$ to approach some finite value as $k$ approaches zero \cite{To18a}. We note that $S(k)$ is on the order of $10^{-9}$ at the smallest $k$ directly accessible in simulations, indicating a very high degree of hyperuniformity for these mesophases resolved by SCFT simulations, while the Quantizer point patterns used to seed the SCFT simulations are only effectively hyperuniform, with their $S(k)$ on the order of $10^{-3}$ at the smallest $k$. This difference in the degree of hyperuniformity suggests that the local relaxation due to the simultaneous minimization of interfacial area between different domains and chain stretching of block copolymers helps drive the system towards hyperuniformity. It is also noteworthy that slight polydispersity develops for the micelles in the DHU state as the result of the SCFT relaxation, which contributes to the perfect hyperuniformity of the final micelle structures as well. For example, the polydisperse degree of the micelles measured by $\gamma=r_{max}/r_{min}$ is around 1.2 at $\chi N = 17$ and $f_A = 0.25$, where $r_{max}$ and $r_{min}$ are the radii of the largest and smallest micelles, respectively. In addition, the average domain spacing between neighboring spheres is around 1.2\% swollen in the DHU state compared to that in the BCC sphere phase, leading to larger distance between neighboring spheres, as indicated by the location of the primary peak in $S(k)$.

\begin{figure}[ht!]
\begin{center}
$\begin{array}{c}\\
\includegraphics[width=0.50\textwidth]{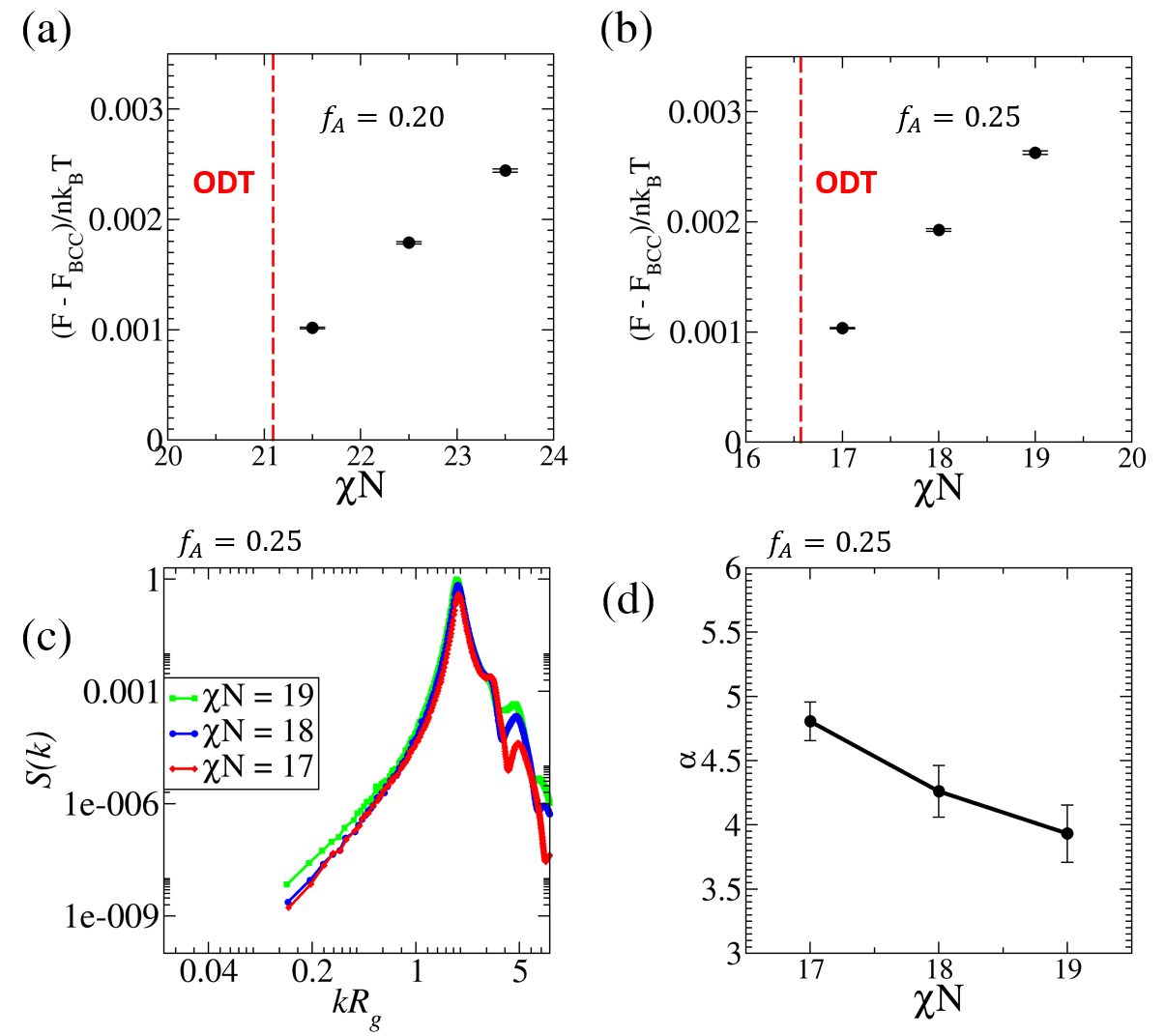} 
\end{array}$
\end{center}
\caption{(Color online) Mean free energy difference between the metastable DHU micelles and the thermodynamically stable BCC sphere mesophase as a function of the segregation strength $\chi N$ at monomer volume fractions $f_A  = 0.20$ (a) and $f_A  = 0.25$ (b). The results are averaged over 10 SCFT solutions, and the error bar of the free energy difference is also shown. The segregation strength $(\chi N)_{ODT}$ at the ODT is shown using red dashed lines for $f_A = 0.20$ and $f_A = 0.25$. Ensemble-averaged $S(k)$ and the corresponding small-$k$ exponent $\alpha$ in $S(k) \sim k^\alpha$ as functions of the segregation strength $\chi N$ at $f_A = 0.25$ are shown in (c) and (d), respectively.} \label{fig_2}
\end{figure}

Next, we investigate the energetics of the DHU mesophases. In Fig. \ref{fig_2}(a) and (b) we show the free energy difference between these metastable phases and the global free energy minimum of BCC spheres as a function of segregation strength $\chi N$ at $f_A = 0.20$ and $f_A = 0.25$. Interestingly, these DHU micelle mesophases possess very similar free energy to the BCC sphere mesophase. For example, at $\chi N = 17$ and $f_A = 0.25$, the mean free energy of the DHU micelles is only $1.0\times 10^{-3} k_BT$ per chain higher than that of the BCC structure, and $6.8\times 10^{-3} k_BT$ per chain lower than the disordered homogeneous state (with $\langle F_{DHU} \rangle \approx 1.0003\times F_{BCC}$), where $k_B$ is the Boltzmann constant. By contrast, the mean free energy of the previously obtained \cite{Do23} disordered fluid-like packings is approximately $2.5\times 10^{-3} k_BT$ per chain higher than that of the BCC morphology, which is considerably higher than that of our DHU mesophases. In other words, DHU micelles are much more favorable energetically than previously obtained liquid-like packings \cite{Do23}. Due to the Quantizer initial conditions and state-of-the-art optimization algorithms that we are using, all the DHU micelle mesophases obtained after SCFT relaxations at a given $\chi N$ and $f_A$ possess very similar free energies from different runs, which is different from the distribution of free energies reported in Ref. \cite{Do23}. For example, the standard error associated with the free energy is only 5.0$\times 10^{-6} k_BT$ per chain at $\chi N = 17$ and $f_A=0.25$. Moreover, we find that the relative stability of DHU micelles increases as the segregation strength decreases and approaches the order-disorder transition (ODT), regardless of the monomer species volume fraction $f_A$. In particular, the free energy difference between DHU micelles and BCC spheres increases approximately linearly as $\chi N$ increases away from the ODT. Here the segregation strength $(\chi N)_{ODT}$ at the ODT is estimated to be 21.09 and 16.57 for $f_A = 0.20$ and $f_A = 0.25$, respectively, from SCFT simulations of the disordered homogeneous phase and the BCC sphere mesophase. As temperature decreases and $\chi N$ becomes sufficiently large [e.g., $\chi N - (\chi N)_{ODT} \gtrsim 5$ at $f_A = 0.20$], the free energy difference between DHU micelles and BCC spheres (scaled by the thermal energy) becomes large enough (close to the free energy difference between BCC spheres and hexagonally-packed cylinders) that will make the formation of DHU mesophases difficult. On the other hand, as temperature increases and $\chi N$ decreases below $(\chi N)_{ODT}$, the system is no longer microphase-separated and becomes trivially homogeneous, which is not interesting for typical applications involving DHU materials. These results indicate that the emergence of disordered hyperuniform mesophases is most likely to be observed in the weak segregation regime. In addition, we also confirm that the hyperuniformity of the disordered micelle mesophases is robust with respect to the system size and compressibility of the block copolymer melts (see SM for details).


As $\chi N$ varies at a given $f_A$, the morphology of the DHU micelle mesophases also changes. As an example, in Fig. \ref{fig_2}(c) we show $S(k)$ at different $\chi N$ for $f_A = 0.25$, which all indicate perfect hyperuniformity for the corresponding micelle mesophases. We then proceed to determine the small-$k$ scaling exponent $\alpha$ in $S(k) \sim k^\alpha$ by looking at the large-$t$ scaling behavior of the excess diffusion spreadability $\mathcal{S}(\infty) - \mathcal{S}(t)$ computed from $S(k)$ (see SM for details) \cite{Wa22}. We find that as $\chi N$ increases, the exponent $\alpha$ decreases as shown in Fig. \ref{fig_2}(d), which can be explained by the fact that as $\chi N$ increases, the A-rich and B-rich regions become richer in A and B, respectively, and the interphase becomes sharper, leading to larger density fluctuations and smaller $\alpha$. The $\alpha$ values in Fig. \ref{fig_2}(d) are also much larger than the scaling exponent $\alpha=2$ predicted by the random phase approximation (RPA). The RPA calculation extracts the linear response of the disordered homogeneous state and is only valid when $\chi N$ is below $(\chi N)_{ODT}$. Furthermore, the mean distance between neighboring spheres in these mesophases increases by about 3.0\% as $\chi N$ increases from 17 to 19 at $f_A = 0.25$, as reflected by the shift of the primary peak of $S(k)$ to smaller $k$, which is accompanied by a proportional increase in the mean size of micelles.

To summarize, in this work we demonstrated the emergence of disordered hyperuniform mesophases in neat melts of block copolymers by simulating large samples containing as many as 2000 micelles. Moreover, we showed that these disordered hyperuniform mesophases derived from point configurations corresponding to the local energy minimum of the Quantizer energy possess very similar free energies to the thermodynamically stable BCC structure and are more favorable energetically than previously studied \cite{Do23} disordered fluid-like packings. Our findings suggest an inexpensive route to make disordered hyperuniform materials given the wide commercial availability of block copolymers, and may lead to new applications. For example, the isotropic nature of the disordered hyperuniform micelle mesophases and their selective scattering of certain wavelengths of electromagnetic waves might enable the design of novel non-iridescent structural colors \cite{Yu21} if the distances between the micelles are tuned to optically relevant length scales. This could be facilitated both through polymer architectural design, e.g. moving from linear to bottlebrush block polymers, as well as by the addition of selective solvents.

It is also noteworthy that while we are unaware of experimental evidence suggesting that the observed liquid-like packings \cite{Ba17, Lo22} are hyperuniform, given the low free energies of the metastable DHU micelles, it is entirely possible that some of those liquid-like packings will be verified to be DHU in the future. The recent development of high-resolution real-space imaging techniques \cite{Ya21b, Ka22, Su23} for soft-matter systems present new opportunities for such experiments, where one could directly compute the structure factor or local variance associated with fluctuations in the field \cite{To16a} from the extracted real-space microstructures and ascertain the hyperuniformity of the mesophases. Moreover, the realization of DHU structures in neat melts of block copolymers might be facilitated by tuning the thermal processing procedures.

DHU micelle mesophases are also similar to the maximally random jamming (MRJ) state \cite{Ma23} and perfect glass state \cite{Zh16b} in that they are all deep free energy minima, and are hyperuniform. However, unlike the MRJ state and perfect glass state that can never crystallize \cite{Zh16b}, our DHU micelle mesophases possess higher free energies than the thermodynamically stable ordered BCC micelle mesophase and the energy barrier is finite, so the DHU micelle mesophases may crystallize over long periods of time. In addition, the DHU micelle mesophases are realizations of random scalar fields, which differ from the cases of the MRJ state and perfect glass state where one is dealing with discrete particles. Such a random field, of course, approximates a (soft) sphere packing but with a slight degree of polydispersity as discussed above, and can take any value between 0 and 1 instead of a binary value (0 or 1) at a given spatial location. These additional degrees of freedom facilitate the formation of hyperuniform materials similar to the top-down tessellation-based procedure for two-phase media from Ref. \cite{Ki19a}. Here, however, we suggest a bottom-up approach.

\begin{acknowledgments}
D. C. and G. H. F. were supported by the U.S. Department of Energy, Office of Basic Energy Sciences, under Award Number DE-SC0019001. This work made use of the BioPACIFIC Materials Innovation Platform computing resources of the National Science Foundation Award No. DMR-1933487. Use was also made of computational facilities purchased with funds from the National Science Foundation (OAC-1925717 and CNS-1725797) and administered by the Center for Scientific Computing (CSC). The CSC is supported by the California NanoSystems Institute and the Materials Research Science and Engineering Center (MRSEC; NSF DMR-2308708) at UC Santa Barbara. M. A. K. acknowledges funding and support by the Deutsche Forschungsgemeinschaft (DFG, German Research Foundation) through the SPP 2265, under grant numbers KL 3391/2-2, WI 5527/1-1 and LO 418/25-1, and by the Initiative and Networking Fund of the Helmholtz Association through the Project ``DataMat''. We also thank Dr. Ge Zhang for supplying us configurations of disordered stealthy hyperuniform point patterns, and Dr. Christopher Balzer and Timothy Quah for helpful discussion.
\end{acknowledgments}

%

\end{document}